\documentclass[10pt]{article}
\usepackage{lipsum}
\usepackage{authblk,etoolbox}

\makeatletter
\patchcmd{\@maketitle}{center}{flushleft}{}{}
\patchcmd{\@maketitle}{center}{flushleft}{}{}

\def\maketitle{{%
  
  \AB@maketitle}}
\makeatother

\usepackage{verbatim}
\usepackage[small,tight,FIGTOPCAP]{subfigure}
\usepackage{amsmath}
\usepackage{amsfonts}
\usepackage{amssymb}
\usepackage[]{mathrsfs}
\usepackage{xcolor}
\usepackage{anysize}
\usepackage{epsfig}
\usepackage{bm}
\usepackage{setspace}
\usepackage{enumerate}
\usepackage{graphicx}
\usepackage{dcolumn}
\usepackage{bm}
\usepackage{enumitem}
\usepackage{citesort}  
\usepackage[superscript]{cite}

\newcommand{\pare}[1]{\left( #1 \right)}

\newcommand{\cor}[1]{\left[ #1 \right]}

\newcommand{\ave}[1]{\left\langle #1 \right\rangle}

\renewcommand{\epsilon}{\varepsilon}

\usepackage[a4paper,left=2.5cm, right=2.5cm, top=2cm, bottom=2.0cm]{geometry}
\usepackage{lipsum}
\newcommand\blfootnote[1]{%
  \begingroup
  \renewcommand\thefootnote{}\footnote{#1}%
  \addtocounter{footnote}{-1}%
  \endgroup
}

\makeatletter
\renewcommand\@biblabel[1]{\textsuperscript{#1}}
\makeatother
\usepackage{setspace}
\begin{document}
\title{\singlespacing{\flushleft{\LARGE{\textbf{Noise-assisted energy transport in electrical oscillator networks with off-diagonal dynamical disorder}}}}}
\author[1,2$\dagger$]{Roberto de J. Le\'{o}n-Montiel}
\author[3,4$\dagger$]{Mario A. Quiroz-Ju\'{a}rez}
\author[3$\dagger$]{Rafael Quintero-Torres}
\author[3]{Jorge L. Dom\'{i}nguez-Ju\'{a}rez}
\author[1]{H\'{e}ctor M. Moya-Cessa}
\author[5,6]{Juan P. Torres}
\author[3]{Jos\'{e} L. Arag\'{o}n}
\affil[1]{\emph{Instituto Nacional de Astrof\'{i}sica, \'{O}ptica
y Electr\'{o}nica, Calle Luis Enrique Erro 1, Santa Mar\'{i}a
Tonantzintla, Puebla CP 72840, M\'{e}xico}}
\affil[2]{\emph{Department of Chemistry \& Biochemistry, University of California San Diego, La Jolla, California 92093, USA}}
\affil[3]{\emph{Centro de F\'{i}sica Aplicada y Tecnolog\'{i}a Avanzada, Universidad
Nacional Aut\'{o}noma de M\'{e}xico campus Juriquilla, Boulevard
Juriquilla 3001, Juriquilla Quer\'{e}taro 76230, M\'{e}xico}}
\affil[4]{\emph{Escuela Superior de Ingenier\'{i}a Mec\'{a}nica y
El\'{e}ctrica, Culhuac\'{a}n. Instituto Polit\'{e}cnico Nacional,
Santa Ana 1000, San Francisco Culhuac\'{a}n 04430, Distrito
Federal, M\'{e}xico.}}
\affil[5]{\emph{ICFO - Institut de Ci\`encies Fot\`oniques, Mediterranean Technology Park, 08860
Castelldefels (Barcelona), Spain}} \affil[6]{\emph{Department of
Signal Theory and Communications, Jordi Girona 1-3, Campus Nord
D3, Universitat Polit\`ecnica de Catalunya, 08034 Barcelona,
Spain}}

\date{}
\maketitle \blfootnote{$\dagger$ These authors contributed equally
to this work.} \vspace{-10mm}

\large\textbf{Noise is generally thought as detrimental for energy
transport in coupled oscillator networks. However, it has been
shown that for certain coherently evolving systems, the presence
of noise can enhance, somehow unexpectedly, their
transport efficiency; a phenomenon called environment-assisted
quantum transport (ENAQT) or dephasing-assisted
transport. Here, we report on the experimental observation of such effect in a network of coupled
electrical oscillators. We demonstrate that by introducing stochastic fluctuations in one of the couplings of the network, a relative enhancement in the energy transport efficiency of $\mathbf{22.5 \pm 3.6\,\%}$ can be
observed.}

Transport phenomena are ubiquitous throughout different fields of
research. Some of the most common examples of transport analysis
are seen in the fields of physics, chemistry and biology
\cite{plawsky_book}. In recent years, energy transport assisted by noise \cite{mohseni2008,plenio2008,caruso2009} has attracted a great deal of attention, partly because of its potential role in the development of future artificial light-harvesting technologies \cite{ball2011,lambert2013,huelga2013}. This intriguing phenomenon has theoretically been shown to occur in several quantum \cite{rebentrost2009,chin2010,caruso2011,mostame2012,shabani2012,kassal2012,roberto2014} and classical \cite{kramer1993,hanggi2009,roberto2013,spiechowicz2014} systems; however, efforts towards its experimental observation had not been presented until very recently. Viciani \emph{et al.} \cite{viciani2015} showed an enhancement in the energy transport of optical fiber cavity networks, where the effect of noise on the system was introduced by averaging the optical response of several network configurations with different cavity-frequency values. In a closely related experiment, Biggerstaff \emph{et al.} \cite{biggerstaff2015} demonstrated an increase in the transport efficiency of a laser-written waveguide network, where decoherence effects were simulated by averaging the output signal of the waveguide array considering different illumination wavelengths. Using the same photonic platform, Caruso \emph{et al.} \cite{caruso2015} observed an enhanced transport efficiency when suppressing interference effects in the transport dynamics of a photonic network. In this experiment, noise was implemented by dynamically modulating the propagation constants of the waveguides, which is the natural way for producing decohering noise, as it has been experimentally demonstrated in the context of quantum random walks \cite{broome2010,schreiber2011,jiri2013}.

In this work, we report on the observation of noise-assisted energy transport in a network of
capacitively coupled $RLC$ oscillators, where $R$ stands for
resistance, $L$ for inductance, and $C$ for capacitance. Although
in previous studies of ENAQT noise has been modeled as
fluctuations in the frequency of each oscillator, so-called diagonal fluctuations \cite{mohseni2008}, here we introduce noise in the system by means of stochastic fluctuations in one of the network's
capacitive couplings, referred to as off-diagonal dynamical disorder \cite{rudavskii2000}. Using this system, we show that fluctuations in the coupling can indeed influence the system so
that the energy transferred to one of the oscillators is
increased, demonstrating that off-diagonal dynamical disorder can effectively be used for
enhancing the efficiency of energy transport systems.

\section*{Results}
We consider a network of three identical $RLC$ oscillators (as
shown in Fig. \ref{fig:circuit}), whose dynamics are described by
\begin{eqnarray}
\frac{d V_{n}}{dt} &=& -\frac{1}{C}\cor{i_{n} + \frac{V_{n}}{R} + \sum_{m\neq n}^{3} C_{nm}\pare{\frac{d V_{n}}{dt} - \frac{d V_{m}}{dt}} }, \label{eq:voltage}\\
\frac{d i_{n}}{dt} &=& \frac{V_{n}}{L}, \label{eq:current}
\end{eqnarray}
where $V_{n}$ is the voltage in each oscillator, $i_{n}$ is the current, and $C_{nm}$ represents the
capacitive couplings.

Noise is introduced in the system by inducing random fluctuations
in one of the capacitive couplings, so that
\begin{equation}
C_{12}\pare{t} = C_{12}\cor{ 1 + \phi\pare{t}},
\label{eq:coupling}
\end{equation}
with $C_{12}$ being the average capacitance of the coupling and
$\phi\pare{t}$ a Gaussian random variable with zero average, i.e.
$\ave{\phi\pare{t}} = 0$, where $\ave{\cdots}$ denotes stochastic
averaging.

Previous studies of noise-assisted transport have shown that
efficiency enhancement can be observed by measuring the energy
that is irreversibly dissipated in one particular site of the
network, the so-called sink or reaction
center\cite{plenio2008,rebentrost2009,roberto2014,roberto2013}.
Here, we take the resistance in Oscillator-2 to be the sink and
measure the relative energy that is dissipated through it by
computing
\begin{equation}\label{eq:efficiency}
W = \frac{1}{\mathcal{N}}\int_{0}^{\infty}\ave{\frac{V_{2}^{2}\pare{t}}{R}}dt,
\end{equation}
where $\mathcal{N}$ is the total energy dissipated
by all the oscillators. Notice that Eq. (\ref{eq:efficiency}) is
equivalent to the efficiency measure that was derived in a
previous work on noise-assisted transport in classical
oscillator systems \cite{roberto2013}.


We have experimentally implemented the system
described by Eqs. (\ref{eq:voltage})--(\ref{eq:coupling}) using
functional blocks synthesized with operational amplifiers and
passive linear electrical components (see Methods and accompanying
Supplementary Information). Our experimental setup was designed so
that the frequencies of each oscillator were the same ($\nu =
290.57$ Hz), as well as the couplings between them ($C_{nm} =
40\;\mu$F). Notwithstanding, based on our measurements, we found
that the designed system was characterized by the following
parameters: $\nu = 283.57$ Hz, $C_{12} = 39\;\mu$F, $C_{13} =
41.7\;\mu$F, and $C_{23} = 39\;\mu$F. These variations in the
parameters of the system result from the tolerances (around $5\%$)
of all the electronic components used in the implementation of the
circuit.

\begin{figure}[h!]
\begin{center}
\raisebox{3mm}{\includegraphics[width=13.0cm]{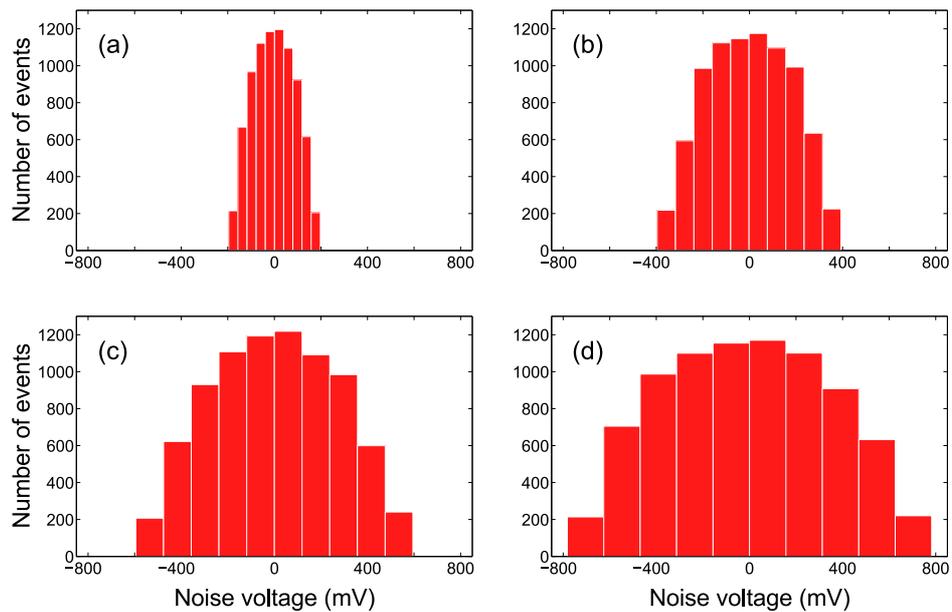}}
\end{center}
\caption{Histograms of noise signals extracted directly from the
arbitrary function generator using different values of noise
voltage: (a) $V_{s} = 200$ mV, (b) $V_{s} = 400$ mV, (c) $V_{s} =
600$ mV and (d) $V_{s} = 800$ mV. Samples were obtained from noise
signals with frequency of 1 kHz, within a 1s time-window. Number
of events is defined as the number of samples that have the same
voltage.} \label{fig:histograms}
\end{figure}

As described in the methods section, noise was introduced in one
of the network's capacitive couplings using a random signal
provided by a function generator (Diligent Analog Discovery
410-244P-KIT). The electronic circuit was designed so the voltage
of the noise signal, $V_s$, is directly map into Eq.
(\ref{eq:coupling}), thus making the stochastic variable
$\phi\pare{t}$ fluctuate within the interval $(-V_s, +V_s)$, with
a 1 kHz frequency. Notice that the frequency of the noise signal is higher than the oscillators' natural frequency, which guarantees a true dynamical variation of the capacitive coupling.  Figure \ref{fig:histograms} shows some examples of the histograms of noise signals extracted directly from the
function generator.

\begin{figure}[h!]
\begin{center}
\raisebox{3mm}{\includegraphics[width=13.0cm]{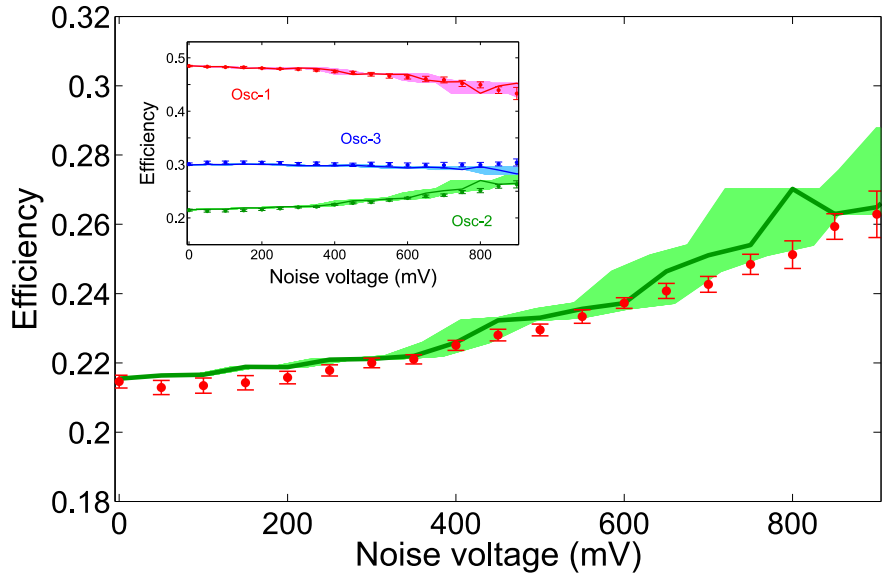}}
\end{center}
\caption{Transport efficiency measured in Oscillator-2 as a function of
the noise voltage. Experimental results (dotted line) were
obtained by averaging the oscillator's signal over 500 stochastic
realizations. The solid line represents the theoretical
calculation of transport enhancement using the noise signal
extracted directly from the arbitrary function generator. The
shaded region represents transport enhancement when noise
introduced in the capacitive coupling deviates from the value
provided by the function generator by up to $10\%$. Transport
efficiencies of all oscillators are plotted in the inset. Notice
that the effect of noise is to rearrange the energy available in
the system in order to increase the efficiency of Oscillator-2. Error
bars correspond to one standard deviation.} \label{fig:eff1}
\end{figure}

Using the configuration described above, we measured the energy
dissipated through the resistor in Oscillator-2 [Eq.
(\ref{eq:efficiency})] as a function of the noise voltage
introduced in the capacitive coupling. We can see from Fig.
\ref{fig:eff1} that transport efficiency in Oscillator-2 is
enhanced as the noise voltage increases, a sign of noise-assisted
energy transport \cite{roberto2013}. We obtained an enhancement of
$22.50\pm 3.59\,\%$ for a maximum noise voltage of $850$ mV. It is
important to remark that we cannot go beyond this value using the
present configuration because, when introducing noise signals
close to $1$ volt (or more), the system becomes unstable due to
the presence of negative capacitances in the coupling [see Eq.
(\ref{eq:coupling})]. However, as obtained in our numerical simulations, a higher efficiency may be reached by
incorporating random fluctuations in the remaining couplings.

To verify that the observed enhancement was a consequence of
energy rearrangement due to random fluctuations in the coupling,
and not because external energy was introduced in the electronic
circuit, we measured the transport efficiency of all oscillators.
We can see from the inset in Fig. \ref{fig:eff1} that transport
efficiency of Oscillator-2 is enhanced only because the energy
dissipated through Oscillator-1 becomes smaller. This clearly
shows that the effect of noise is to create new pathways in the
system through which energy can efficiently flow towards an
specific site, generally referred to as sink or reaction center.

\section*{Discussion}

The results presented here show that noise-assisted transport, an ubiquitous concept that may help us understand efficient energy transport in diverse classical and quantum systems, can be observed in simple electronic circuit networks. This opens fascinating routes towards new methods for enhancing the efficiency of different energy transport systems, from small-scale RF and microwave electronic circuits to long-distance high-voltage electrical lines. In this way, a specific feature initially conceived in a quantum scenario (environment-assisted quantum transport) has shown to apply as well in classical systems, widening thus the scope of possible quantum-inspired technological applications.

\section*{Methods}

\begin{figure}[h!]
\begin{center}
 \includegraphics[width=15.75cm]{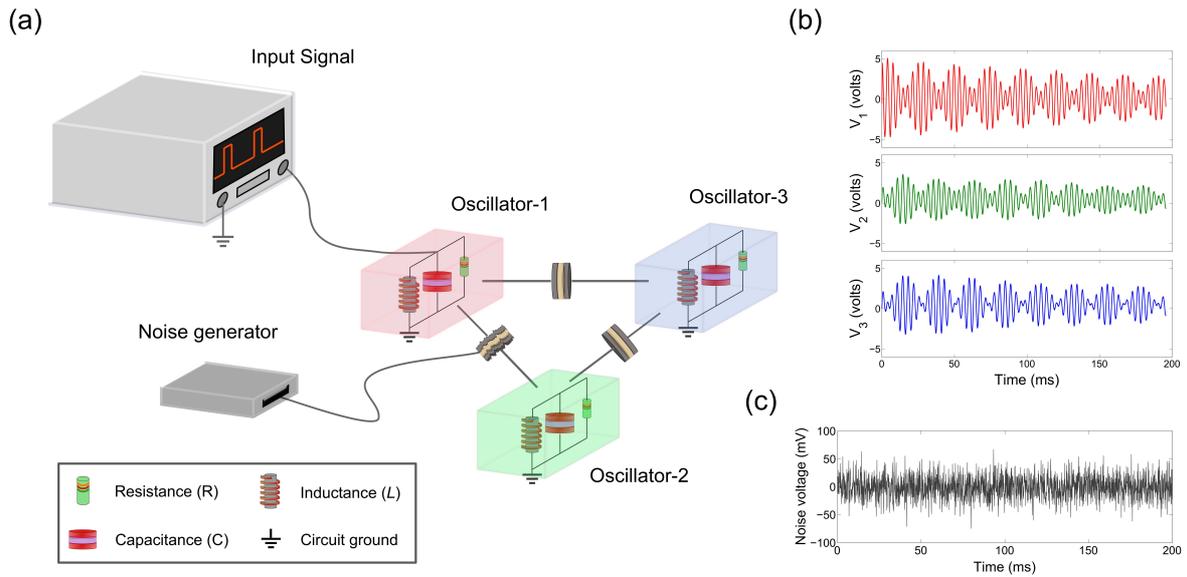}
\end{center}
\caption{Schematic representation of the experiment and typical
dynamic interaction response. (a) Schematic of the network of
three capacitively coupled $RLC$ electrical oscillators. (b)
Typical averaged voltage signals measured in each oscillator. (c)
Sample of a typical noise signal, $V_{s}$, introduced in the
capacitive coupling between the first and second oscillator.}
\label{fig:circuit}
\end{figure}

In our experiment, three identical \emph{RLC} electrical
oscillators interact by means of three ideal
capacitors. The input energy is injected into a
single oscillator (Oscillator-1), and we follow the dynamics of
the system by registering independently the voltage of each
oscillator (Fig. \ref{fig:circuit}). Noise is introduced in the
system by means of random fluctuations in one of the capacitive
couplings, where the magnitude of the changes in the capacitance
is defined by the noise voltage provided by an arbitrary function
generator (Diligent Analog Discovery 410-244P). Relative changes
in the capacitance range from 0\% (no voltage) to 100\% (1 V) in a
controllable way. To avoid instabilities, the
1-KHz-frequency noise signal was varied from 0 to 850 mV, which
corresponds to a range of the fluctuating capacitance going from
$C_{12}$ to $(1 \pm 0.85)C_{12}$.

The voltage of each oscillator contains the information about the
stored energy in the system as well as the energy dissipated by
each oscillator. A Tektronix MSO4034 oscilloscope (impedance
1M$\Omega$) is used to measure these voltages. The voltage signals
were extracted from the oscilloscope using a PC-OSCILLOSCOPE
interface, which transfers the information through a USB port.
Because we are working with stochastic events, measurements were
repeated up to 500 times for each noise voltage, and averaged
using a MatLab script. The initial input signal
is a single pulse, with a pulse duration of 200 $\mu$s, injected
in Oscillator-1 using an Arbitrary Waveform Generator from Agilent
33220A, with a 5 Hz frequency. The high level voltage amplitude is
5 V, while the low-level voltage amplitude is 0 V.

The synthesized transformation of the electronic circuit to an
analog computer is described in the Supplementary Information.
Basically, the analog computer makes analog simulations of
differential equations using electronic components. The input and
output voltages of an electronic circuit correspond to
mathematical variables. These voltage variables are therefore
representations of the physical variables used in the mathematical
model.

An important issue that we would like to point out is that in this
work both, the initial conditions and noise, are physically
introduced via a voltage signal. This is particularly relevant
because in the first case this voltage represents the initial
current and, in the second case, the statistical distribution of
the noise voltage can be defined independently from the circuit,
thus it is not necessary to produce fluctuations in the physical
properties of the electrical components--namely resistors,
capacitors or inductors--which generally represents a major
challenge \cite{roberto2014_2}.

Using the building blocks described in the Supplementary
Information, it is possible to synthesize current variables and
inductor elements with only resistors, capacitors and voltages. In
our experiment, the components employed for the implementation of
the corresponding building blocks include metal resistors (1\%
tolerance), polyester capacitors and general-purpose operational
amplifiers (MC1458). A DC power source (BK Precision 1761) was
used to generate a $\pm$12 V bias voltage for the operational
amplifiers. The electronic components of each oscillator were mounted and soldered
on a drilled phenolic board (7.5$\times$4.5 cm) to avoid poor
contacts.

\newpage
\section*{References}
\vspace{-1cm}
\renewcommand{\refname}{\large{}}

\vspace{0.75cm}
\section*{Acknowledgments}

RJLM acknowledges postdoctoral financial support from the University of California Institute for Mexico and the United States (UC MEXUS). MAQJ acknowledges CONACyT for a PhD scholarship. RQT thanks DGAPA-UNAM for financial support through grant IN111614. JLDJ thanks Catedras CONACYT-UNAM. JPT acknowledges support from the Severo Ochoa program (Government of Spain), from the ICREA Academia program (ICREA, Generalitat de Catalunya) and from Fundacio Privada Cellex, Barcelona. JLA wishes to thank CONACYT and DGAPA-UNAM for financial support through grants 167244 and IN106115, respectively.

\vspace{0.75cm}
\section*{Author Contributions}

RJLM, MAQJ, RQT, JLDJ, JPT and JLA conceived, designed and
implemented the experimental setup. RJLM and HMMC provided the
theoretical analysis. All authors contributed extensively to the
planning, discussion and writing up of this work.

\vspace{0.75cm}
\section*{Competing Financial Interests}

The authors declare no competing financial interests.

\newpage

\newpage

\newpage

\textcolor{white}{.}
\vspace{1cm}
\begin{flushleft}
\singlespacing{\LARGE{\textbf{Supplementary Information:
Noise-assisted energy transport in electrical oscillator networks with off-diagonal dynamical disorder}}}
\end{flushleft}
\vspace{0.5cm}

\large From an experimental point of view, it is challenging to
implement a set of interacting $RLC$ electrical
oscillators where certain important defining
parameters should change dynamically (noise), even more if the
time needed to visualize the dynamics and interactions is large
compared with the time-constant imposed by the parasitic
attenuation of a real system. In this work we are interested in
noise, its implementation, its effects and how to
control it. In the first instance, one could
attempt to build noisy $RLC$ oscillators using passive components
\cite{roberto2014_2}, but this strategy is in general limited by
the large damping coefficient in the inductor, which produces a
fast energy extinction, thus hindering the
observation of the sought-after effects.

\begin{figure}[h!]
\begin{center}
  \includegraphics[width=13.0cm]{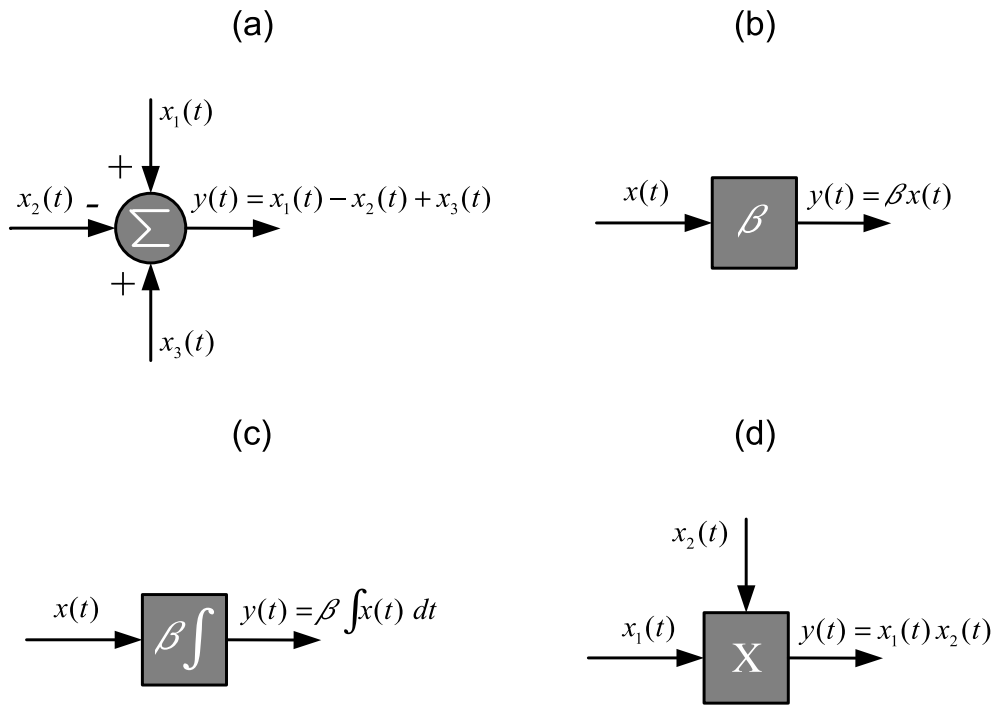}
\end{center}
\caption{Functional Blocks: (a) adder, (b) multiplication by a
  constant, (c) integrator and (d) multiplication of signals.}
\label{fig:Funtional Block}
\end{figure}

In this supplementary information material, we
will show how a network of noisy $RLC$ oscillators can be
electronically implemented by using the working principle of an
analog computer \cite{1S,2S}. Basically, an analog computer makes
analog simulations of differential equations using electronic
components. The input and output voltage variables of an analog
electronic circuit (analog computer) represent physical variables
of the mathematical model that is being studied. In this way, in
order to simulate any mathematical model in an analog computer,
the sequence of mathematical operations involved in the process
must be described by means of functional block diagrams. The four
functional blocks shown in Fig. \ref{fig:Funtional Block} are the
ones used in our experimental setup to implement the mathematical
operations described by Eqs. (1)-(3) of the main
manuscript.

\begin{figure}[h!]
\begin{center}
\includegraphics[width=16.0cm]{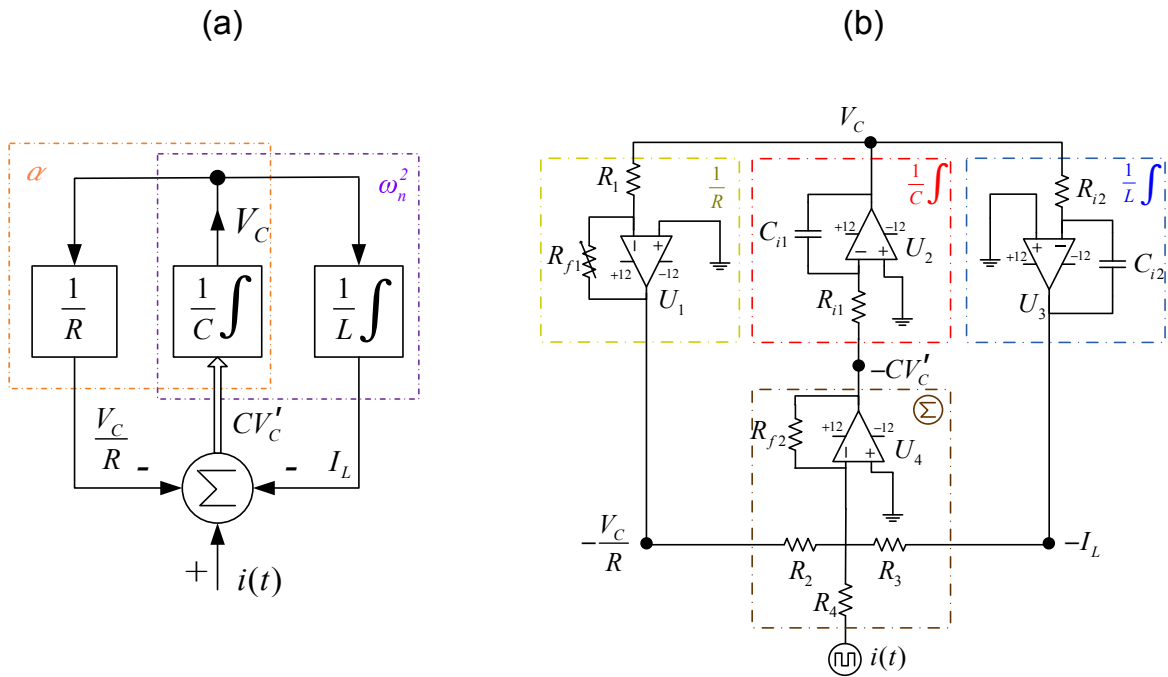}
\end{center}
\caption{Simple $RLC$ parallel oscillator (a) Block diagram and (b)
  Analog electronic circuit.}
\label{fig:Simulation-diagram}
\end{figure}

Using these functional blocks a complete diagram of the oscillator
network can be obtained. Firstly, we build a block diagram for a
simple $RLC$ oscillator [as shown in Fig.
\ref{fig:Simulation-diagram}(a)]. Notice that this diagram
corresponds to the synthesis of a typical damped harmonic
oscillator equation, with $\alpha$ and $\omega$ representing its
damping coefficient and frequency, respectively. Once the block
diagram is built, each mathematical operation can be
electronically implemented with passive linear electrical
components and basic inverting configurations of operational
amplifiers (OPAMPs).

Figure \ref{fig:Simulation-diagram}(b) shows the synthesized
analog circuit for a typical $RLC$ electrical oscillator. Notice
that the inherent sign inversion is a result of the negative
voltage gain of the OPAMP, which must be taken into account in the
design. Here, $R_{j}$, $C_{j}$ and $U_{j}$ stand for resistors and
capacitors, and general-purpose operational amplifiers,
respectively. The values of $R_{f1},R_{1},R_{i1},C_{i1},R_{i2}$
and $C_{i2}$ are defined by the oscillator parameters, that is,
the resistance ($R$), the inductance ($L$) and the capacitance
($C$). They must satisfy:

\begin{equation}
 \label{eq:relation1}
\frac{R_{f1}}{R_{1}} = \frac{1}{R}, \qquad \frac{1}{R_{i1}C_{i1}} =
\frac{1}{C}, \qquad \frac{1}{R_{i2}C_{i2}} = \frac{1}{L},
\end{equation}
and
\begin{equation}
 \label{eq:relation2}
\frac{R_{f2}}{R_{2}} = \frac{R_{f2}}{R_{3}} = \frac{R_{f2}}{R_{4}}=1 .
\end{equation}

This last expression [Eq. (2)] implies that the adder
$\left(U_{4}\right)$ has a unitary amplification factor.

For the experimental setup, we calculated the resistors and
capacitors involved in Fig. \ref{fig:Simulation-diagram}(b)
considering the following parameters: $L=1$ mH, $C=300$ $\mu$F and
$R = 1$ k$\Omega$. The combination of these parameters results in
an oscillation frequency of $f=290.57$ Hz, with a damping
coefficient $\alpha=3.333$ Hz. To design this configuration we
thus take $R_{i1} = 3$ k$\Omega$, $C_{i1}=100$ nF, $R_{i2}=10$
K$\Omega$, $C_{i2}=100$ nF, $R_{1}=1$ k$\Omega$ and $R_{f1} = 3.3$ k$\Omega$.
By selecting $R_{f2}=R_{2}=R_{3}=R_{4}=10$ k$\Omega$, the condition given by
Eq. (2) is satisfied.

To understand how random fluctuations in the coupling between
Oscillator-1 and Oscillator-2 are introduced, we first substitute
Eq. (3) into Eqs. (1) and (2) of the main manuscript, so we obtain
a set of equations given by

\begin{eqnarray}
CV_{1}^{'} &=& -i_{1} - \frac{V_{1}}{R} - C_{x}(V_{1}^{'} - V_{2}^{'})
  (1 + \phi) - C_{x}V_{1}^{'} + C_{x}V_{3}^{'} ,\\
Li_{1}^{'} &=& V_{1} ,\\
CV_{2}^{'} &=& -i_{2} - \frac{V_{2}}{R} + C_{x} (V_{1}^{'} - V_{2}^{'})
  (1 + \phi) - C_{x}V_{2}^{'} + C_{x}V_{3}^{'} ,\\
Li_{2}^{'} &=& V_{2} ,\\
CV_{3}^{'} &=& -i_{3} - \frac{V_{3}}{R} - 2 C_{x} V_{3}^{'} + C_{x}
  V_{1}^{'} + C_{x}V_{2}^{'} ,\\
Li_{3}^{'} &=& V_{3} ,
\label{eq:noiseexpand}
\end{eqnarray}
where we have assumed that the coupling capacitors have the same value, that is, $C_{x}=C_{12}=C_{13}=C_{23}$.

\begin{figure}[t!]
\begin{center}
\includegraphics[width=14.5cm]{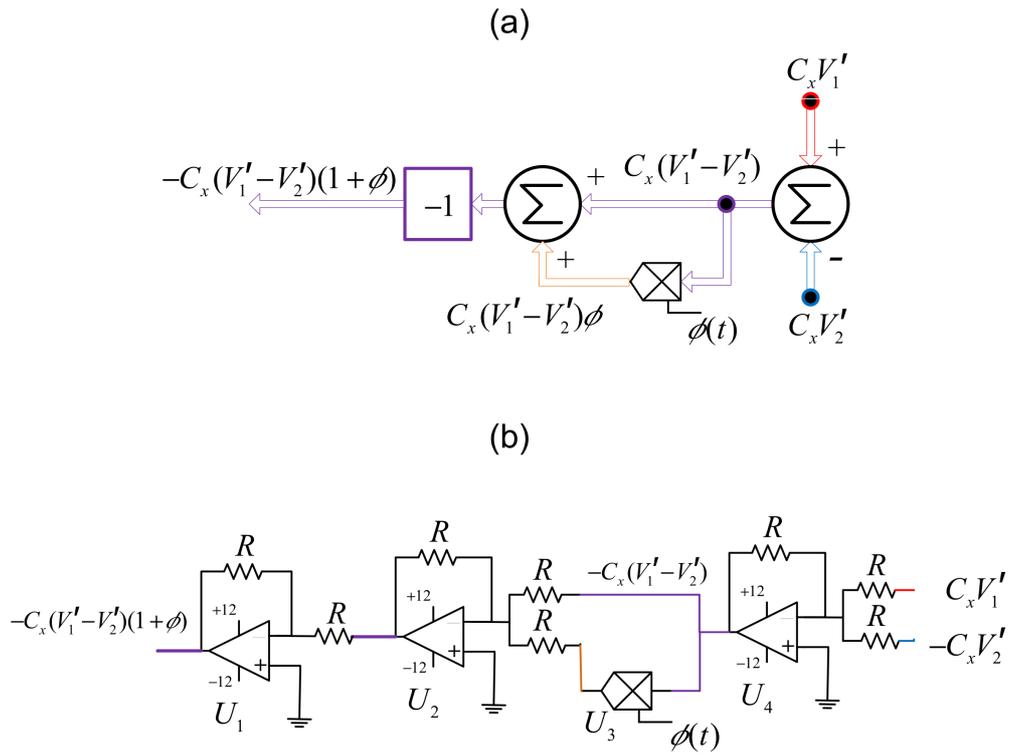}
\end{center}
\caption{Noisy coupling between Oscillator-1 and Oscillator-2: (a) Block
  diagram and (b) Analog electronic circuit.}
\label{fig:BlockNOISE}
\end{figure}

The synthesis of Eqs. (3)-(8) is divided into three parts.
In the first part, the $RLC$ oscillators are implemented using the
electronic circuit in Fig. \ref{fig:Simulation-diagram}. Each
oscillator can be easily implemented using the same electronic
circuit that was obtained for the simple $RLC$ parallel oscillator.
In the second part, the couplings without noise are built using
the same methodology, that is, by designing the block diagram and
then replacing each block with electronic configurations based on
operational amplifiers. Finally, the third part is the design of
the noisy coupling. Note that $CV_{1}^{'}$ and $CV_{2}^{'}$ have
the same noisy coupling term thus it is enough to build only one
of them. Figures \ref{fig:BlockNOISE}(a)-(b) show
the functional block diagram of the noisy coupling, as well as its
equivalent analog electronic circuit.

To conclude, it is important to point out that, in our system,
noise is physically introduced via a voltage signal. This is
particularly relevant because its statistical distribution will be
defined independently from the circuit, which implies that
fluctuations in the physical properties of electronic
components--such as resistors, capacitors or inductors--are not
required. \vspace{1.0cm}

\begin{center}
\large{\textbf{REFERENCES}}
\end{center}
\vspace{-1.5cm}
\renewcommand{\refname}{\large{}}

\newpage

\newpage

\textcolor{white}{.}

\end{document}